# Study of Electromagnetic Properties of Pencil Drawn graphite composite Films on Paper.

Amit R. Morarka*, Aditee C. Joshi
Department of Electronic Science, Savitribai Phule Pune University. Pune-411 007. India
*E-mail: amitmorarka@gmail.com , amitm@electronics.unipune.ac.in

**Abstract**
Graphite has been one of the promising materials in diverse application domains owing to its high conductivity, tunability into different structures and mechanical strength its. The effectiveness of graphite and its derivatives has been studied for electromagnetic domains as well. Pencil strokes on paper create a film of graphite composites which is reported to be useful for fabrication of electronic components. In our study, we extend use of pencil traces on paper for studying its electromagnetic properties. The pencil traces on paper is facile method of coating graphite composite films with relatively lower cost and ease of processing. The interaction of electromagnetic wave with graphite composites produces in modulation of the incident RF power. The RF power was observed to get attenuated with pencil coating on paper as compared to plain paper. The attenuation increased with increasing the signal frequency. Further, stacking more pencil coated papers onto each other results in increasing attenuation factor. Additionally, these pencil coated paper roll was able to attenuate the incoming noise signals in the radio signal reception. This demonstrates potential ability of pencil coated papers to be used for small RF power attenuation applications.

**Keywords**: Graphite, Pencil traces, RF power attenuation,

## 1.0 Introduction

Graphite is one of the most potential materials due to its versatile properties in electrical, thermal and mechanical domains. Additionally graphite can be tailored into various forms like exfoliated graphite, colloidal graphite, flexible graphite, that has been explored in many applications [1-4]. It possesses good electrical conductivity, thermal conductivity; the higher conductivity in graphite can be attributed to availability of delocalized electrons for conduction. The conductivity values are comparable to that of metals. This electrical characteristic of graphite makes it useful candidate in applications for electromagnetism studies. As a result, much of research has been carried out to explore various applications of graphite and its different forms. Flexible graphite sheets have been reported to attenuate the electromagnetic radiation with a power attenuation value of 125-130 dB [4-7]. Furthermore graphite composites prepared with epoxy and polymers have proved to be useful in EMI shielding and microwave absorption properties [2-3].Amongst all the sources and structures of graphite, pencil lead happens to be one of the simplest and low cost source of graphite. Pencil, a routinely used writing tool contains graphite composites in its lead. It consists of graphite added together with intercalated clay and small amount of wax [8-10]. Pencil strokes on surface like paper yield a film of graphite composites. Such kind of Pencil drawn films on paper have yielded many applications including fabrication of passive components like resistor, capacitor and field effect transistors [11].However to the best of our knowledge electromagnetic properties of pencil drawn films on paper have not been explored.

In our study we have used pencil traces to make a film of graphite composites on paper sheets. For this purpose we have coated normal printing papers by putting pencil strokes. The pencil coated papers were characterized for structural characteristics of graphite composites and further investigated for RF power transmission and reception characteristics between the frequencies range 500 MHz-2.5 GHz. The RF power modulation was significant with the addition of pencil coated paper. Further, RF power variation got remarkably improved as number of layers of pencil coated papers was added. Through this study we report pencil traces as very facile and novel technique for modulating incident RF power.

## *2.0 Material and Methods*

### *2.1 Instrument Details*

All the monopole antenna characteristic spectrum was recorded on vector network analyzer by Agilent Technologies E5062A with range of 300 KHz-3GHz. RF generator from Agilent Technologies, N9310A having range of 9kHz-3GHz was used to generate signal of a specific frequency and power. The transmitted power was measured by using a power meter 437B using 8481A power sensor from Hewlett-Packard. Raman spectrum was recorded using Renishaw Raman spectrometer and laser source with excitation wavelength 532 nm was used. Micro balance from Citizen (CX-165) has been used for weighing the plain papers and pencil coated papers.

### *2. 2 Experimental Procedure*

The substrate materials used were readily available printing paper (average of 10 samples , thickness measured using micrometer screw gauge-99.2 μm). The pencils used are 9B grade. The paper sheet is cut into a size of dimensions 21 cm x 7 cm. The complete area of paper was covered with pencil strokes to lead a conductive surface. The effective thickness of the coated graphite was measured by using gravimetric method. For this purpose the weight of the blank paper was recorded before coating and again after coating with pencil. By using the weight difference and other constants thickness of the coated graphite was calculated.

Monopole antennae were fabricated for five different frequencies in range of 500 MHz-2.5 GHz. For each frequency two antennae were fabricated; one is for transmitting the RF power and other as a receiver. Antenna responses were measured by using vector network analyzer. The characteristic spectrum for each antenna is given in supplementary information.

For measuring the RF power attenuation characteristics a setup consisting of two monopole antennae, a mount for holding the pencil coated paper along with RF generator and power meter was used. Monopole antenna was used as a transmitter and a receiver as well. Each antenna was mounted on wooden stand and pencil coated paper was inserted in a mount made up of cardboard on a foam base. In all set up care was taken to ensure that all the mounts and stands were non-conducting in nature. For each frequency RF power is transmitted through the screen and received power was measured using power meter using another monopole antenna placed in front of the mount. The schematic of the setup is given in the following Fig.1a. The photograph of actual setup is also given in Fig.1b. Additionally, setup was slightly changed for measuring the reflected power from graphite coated surface. In this, transmitting and receiving antennae were placed on same side of graphite coated paper. The photograph of the setup is shown in fig. 1c.

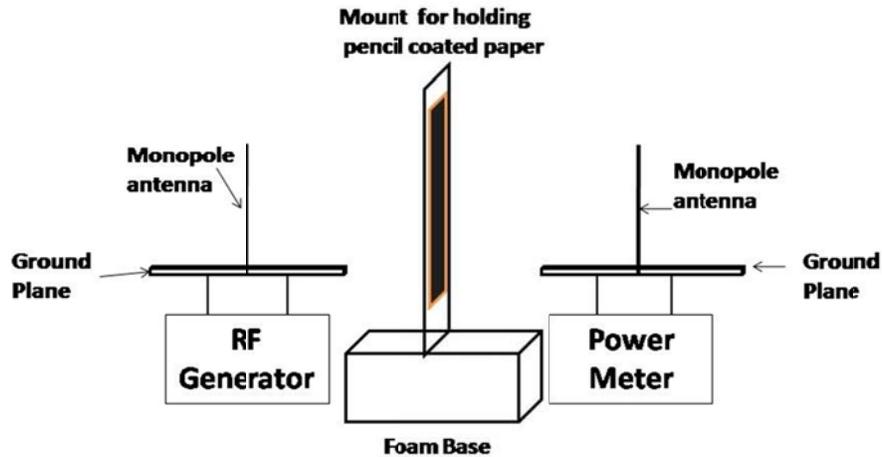

Fig.1a Schematic of the setup used for RF power attenuation.

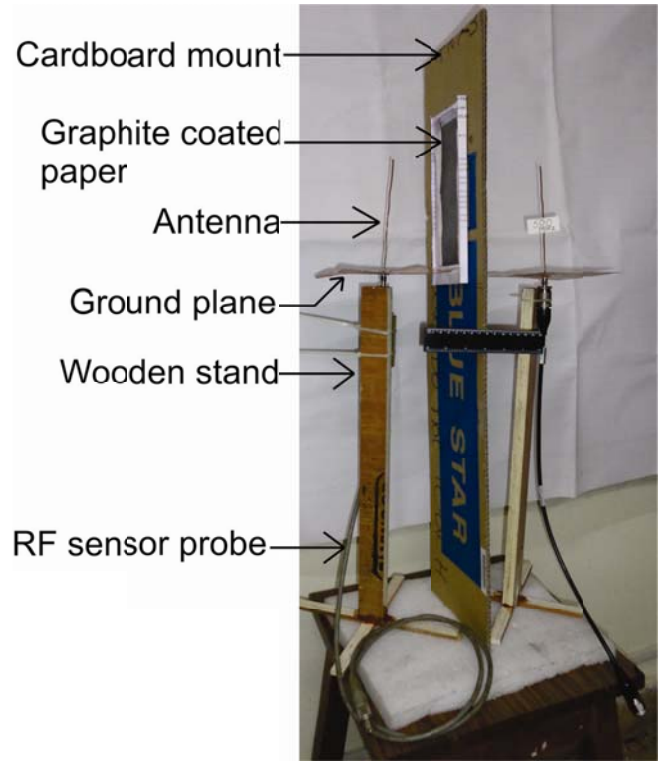

Fig.1b Actual photograph of the setup used for RF power attenuation.

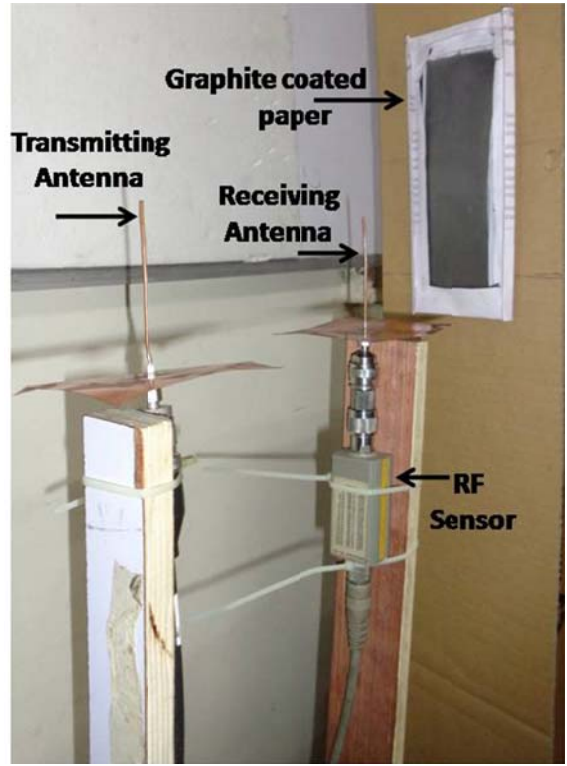

Fig.1c Actual photograph of the setup used for measuring reflected power from graphite surface.

The observed value of antenna frequency was slightly different than the designed value. Antenna dimensions were slightly different than actual designed wavelength hence there was a deviation in frequency. The actual frequency value was set on RF generator such that maximum power was received on the power meter during measurement. In order to study the effect of the pencil coated paper, initially the measurements were made on blank paper without any pencil coatings. Further single pencil coated paper sheet was inserted into the mount and power measurements were performed. Subsequently, other paper sheet was added over previous paper and readings were repeated. In this way total ten pencil coated papers were added onto each other and RF power variation characteristics was studied.

RF power attenuation was measured by taking logarithmic ratio of the received power to incident power. it is given as follows:

$$\text{Attenuation} = 10 \log (P_{received}/ P_{transmitted})$$

## *3.0 Results and Discussion*

The Raman spectrum recorded on graphite composite powder is shown in Fig.2. The observed spectra showed three prominent peaks at 1346 cm$^{-1}$, 1581 cm$^{-1}$ and 2720 cm$^{-1}$ which correspond to the D, G and 2D bands, respectively. This result is in accordance with the Raman spectrum obtained for pencil trace in previous report [11]. The G band arises from stretching of the sp$^2$ bonded carbon lattice and D band originates from the presence of defects in the form of edges and grain boundaries

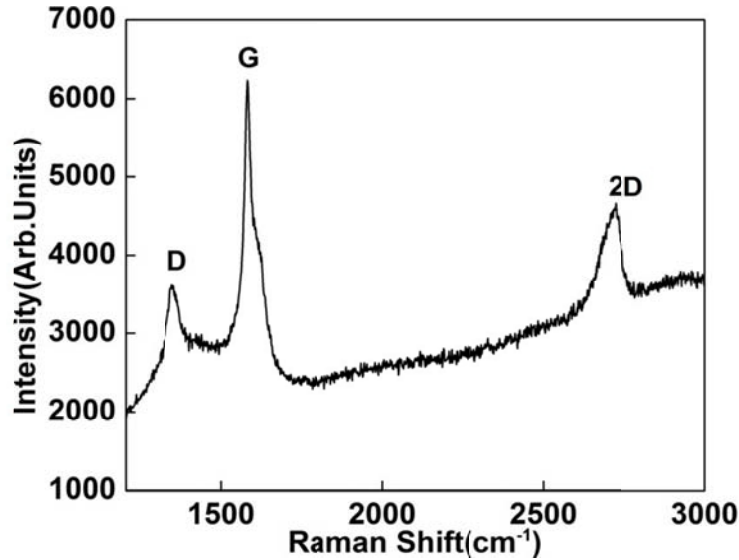

Fig.2 Raman Spectra of graphite composite powder.

The average thickness of graphite composite for all ten papers was ≈ 1.8 μm. The conductivity for the graphite composite was calculated using value of resistance of film and other dimensions such as length and area of paper on which coatings were made. The conductivity value was 457.67 S/m.

RF power response was monitored for blank paper and it was considered as a reference power. Then for each added paper power was recorded and readings were repeated for ten times and average value of power was calculated. This experiment was repeated for all the five frequencies in the range 500 MHz-2.5 GHz. RF power attenuation was calculated for each paper based on average transmitted power. The average attenuation value with addition of number of papers is given in the following Fig.3. The graph shows attenuation characteristic measured at 533 MHz. It can be observed from the graph that addition of first pencil coated sheet absorbs significant amount of RF power that results in attenuation of RF power at the receiver end. The response shows that RF power decays as we add number of pencil coated papers which is similar to previous reports [4-5].

For all the other frequencies similar trend of power decay was observed. As seen in Fig. 3, it can be noted that as we increase the frequency the characteristic attenuation also increased.

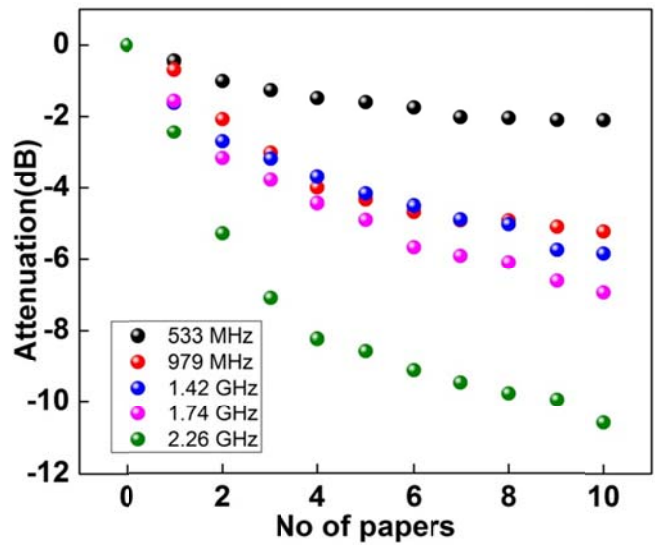

Fig.3 RF power attenuation at different frequencies with total number of papers added.

The RF power attenuation through all ten papers at different frequencies is plotted in the Fig.4. It can be depicted that as frequency increases RF power attenuation also increases. This indicates that out of the amount of power transmitted, substantial amount of power is getting dissipated due to pencil coated paper.

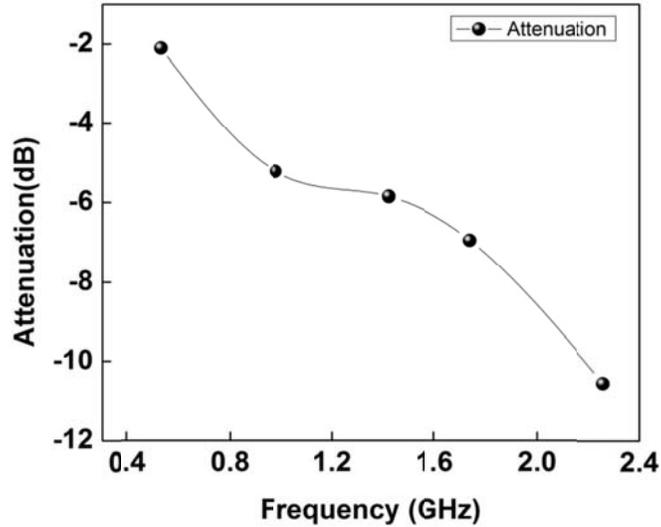

Fig.4 RF power attenuation at different frequencies for all ten papers added.

The RF power attenuation through a specific material depends on thickness of material, skin depth at certain frequency and conductivity. In our study we keep increasing number of papers on the screen effectively increasing the thickness of graphite composites which helps in increasing amount of attenuation for incident RF power.

We have observed that RF signal gets attenuated while propagating through graphite coated papers. Furthermore, we have studied the reflection characteristics of RF signal from graphite coated papers. As shown in fig.1c we have kept transmitting and receiving antenna at same side of graphite coated papers. The distance between the receiver antenna and reflecting surface was varied and RF power was measured. If there is no reflection the power will go on decreasing with increasing distance. Now during distance variations if there is any reflected wave it will interfere with incident wave. As per the standing wave theory [12], this will result in minimum and maximum power at different points. The following graph in figure 5a shows observed power variation for 1.5 GHz frequency. Up to 40 cm distance the power gets reduced with distance but after that power was increased (inset fig. 5a). At this distance we have removed each pencil coated paper and measured power for remaining papers as indicated by Fig. 5b. It can be observed that as we decrease number of graphite coated papers the power goes on decreasing indicative of decrease in reflected component. This shows that graphite coated surface acts like a metallic reflecting surface. We have observed the reflected power in presence of graphite and without any graphite coated paper. Fig.5c shows difference in power for graphite coated papers and without any graphite coated papers at the same distance. For higher frequencies the difference in power with graphite and without graphite surface gets smaller indicative of reduced reflected power. The same trend is observed for different distance values as well, this shows decrease in reflection power with increasing frequency values.

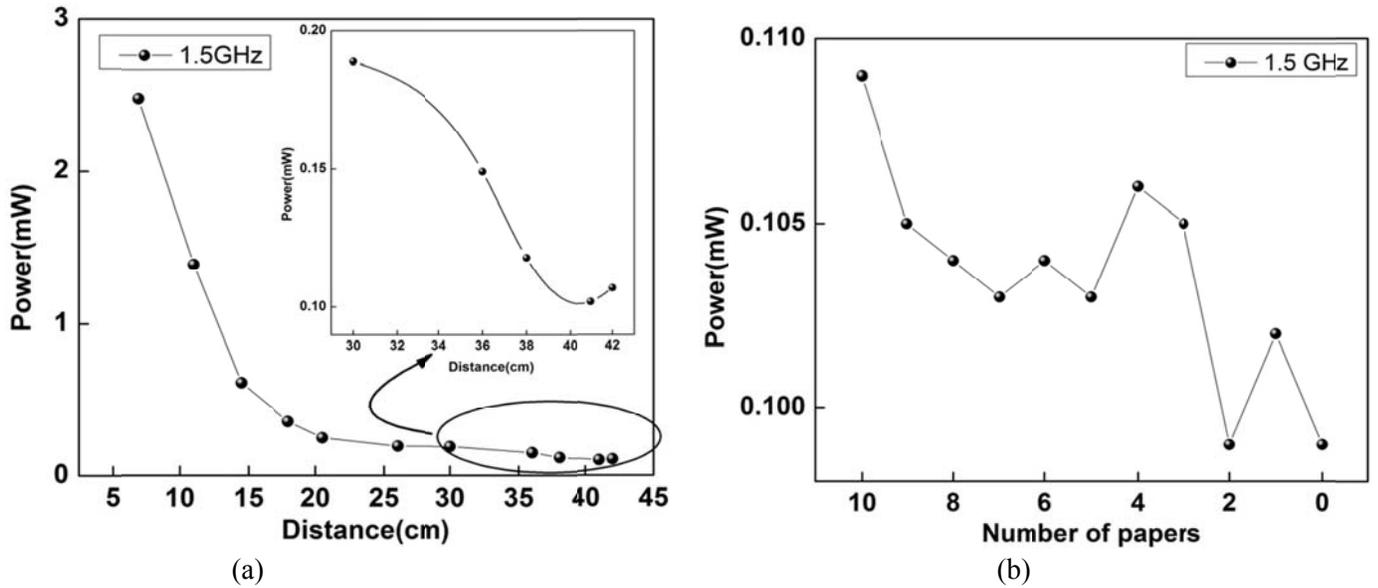

(a)  (b)

Fig.5 a)RF power attenuation 1.5 GHz for all ten papers added. Inset shows increase in power at 40 cm distance. b) Attenuation in reflected power at 40 cm distance with decrease in number of graphite coated papers.

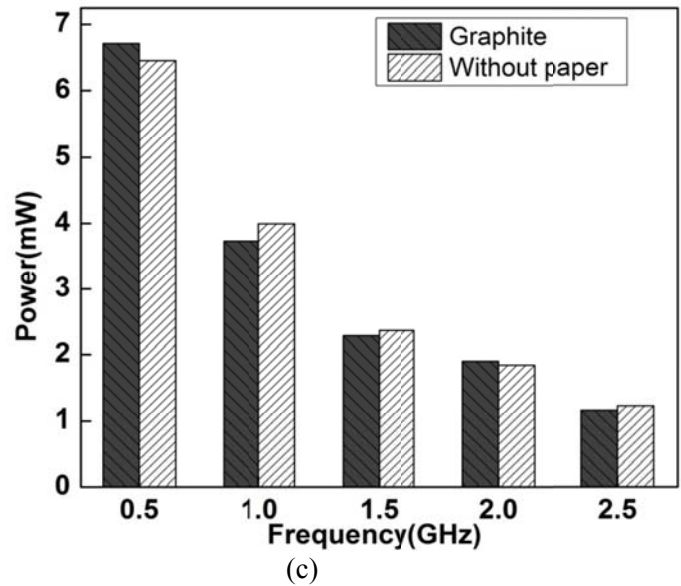

(c)

Fig.5 c) Change in observed at same distance for graphite coated papers and without papers .

### 3.1 Mechanism of RF signal attenuation through pencil coated papers

An electromagnetic wave propagating through a medium undergoes change in its power. While interacting with medium part of wave is reflected, absorbed and remaining component is transmitted. For our system we have studied interaction of electromagnetic wave with graphite coated papers. We have observed the transmitted power and reflected power for variation of frequencies. For transmitted characteristics the attenuation increases with frequency and reflection reduces over the frequency range. This characteristic is similar like metal where high frequency reflection loss gets reduced. The mechanism for this may be envisaged on the basis of properties of graphite composite.

It is known that [13] when an electromagnetic wave propagates through a medium it interacts with the surface charges. The electric field interaction with charges results in certain amount of work done on charges, this result in reduction in energy of wave. Applying this concept, when the electromagnetic wave propagates through the pencil coated paper it interacts with the surface charges present in the graphite composite. Electric field in electromagnetic wave does work on charges present within the volume. This results in utilization of energy of electromagnetic wave. Therefore the energy of electromagnetic wave decreases as work is done on charges. This is indicated by attenuation in RF power at the receiver output after propagating through first pencil coated layer.

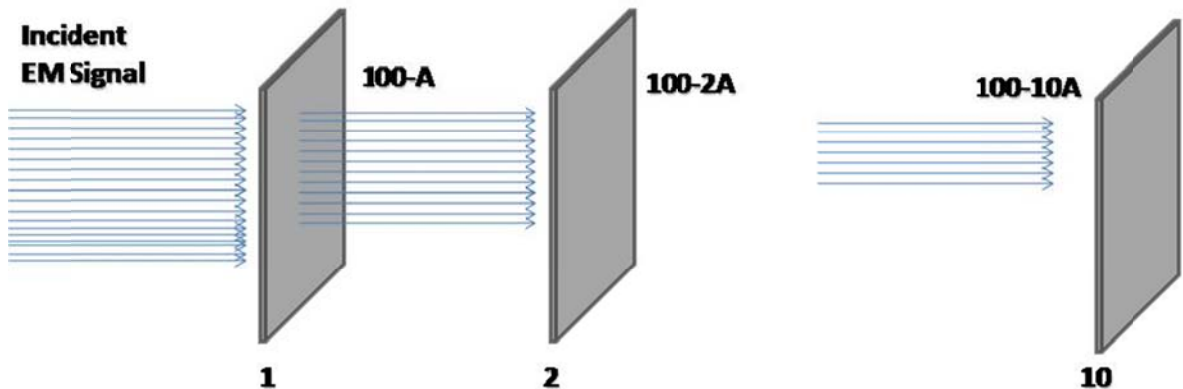

Fig. 6 Illustration depicting probable signal transmission through a stack of ten pencil coated papers.

Furthermore, in our system we have sequentially added number of pencil coated layers adjacent to each other. The expanded view of the system is illustrated in the Fig.6. It shows incident electromagnetic signal propagating through a stack of pencil coated papers. For simplicity we have showed only two papers but system extends for stack of total ten papers.

When the incident signal enters through the first pencil coated paper the signal amplitude is attenuated. Here we consider signal attenuation 'A' at each stage. The incident signal from source can be considered as reference signal 100%. The output signal after passing through first pencil coated paper will be 100-A. It must be noted that the expressions used in the illustration are only for illustrative purpose. Now this attenuated signal acts as incident signal for second pencil coated paper which gets decreased to a value of 100-2A. This means that for the adjacent paper incident signal is attenuated signal from previous stage as depicted in figure. Therefore, as we stack number of papers signal strength is decreased as compared to original signal. The attenuation is more or less similar at each stage. But due to reduction in incident signal amplitude over cascaded stages it appears like attenuated signal amplitude is less than that of first paper. But it is noteworthy that the signal strength itself is less than the original signal so for the same attenuation factor the change may not seem cognizable for a single paper. However in case of stack of ten papers the total attenuation is an integral effect of attenuation at each stage.

The attenuation effect was observed for higher frequencies as well. When the incident electromagnetic wave passes through the first pencil coated paper, the signal attenuation value for lower frequency was 10-15% while as we go for highest frequency attenuation is almost equal to 45%. It is known phenomena that at high frequencies inductive and capacitive reactances govern over resistive elements. Moreover characteristic impedance of graphite increases with increasing frequency [12]. This results in increased resistances leading to increased power dissipation within the material itself. Therefore attenuation increases significantly as compared to the lower frequencies.

### *3.2 Application of pencil coated papers in RF noise attenuation.*

In order to demonstrate the signal attenuation capacity of graphite coated papers, a small radio receiver and a control circuit was used as shown in the following Fig.7. The basic principle underlying the whole unit is whenever radio receives a particular signal the LED connected at the output will be switched on depending on the strength of received signal. The details of the control circuit are added in the supplementary information.

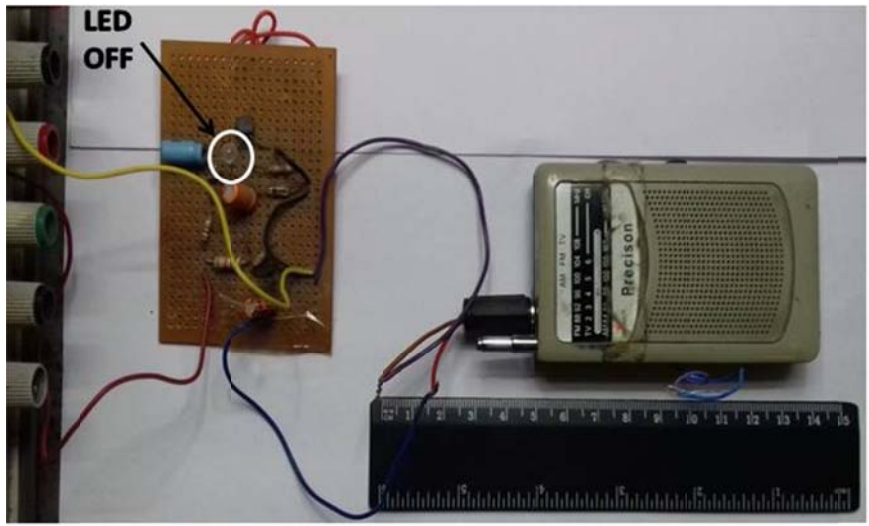

Fig. 7 Radio receiver unit connected to control circuit.

In this circuit whenever there is interference in radio signal due to any peripheral sources, LED intensity will be modulated or LED will keep blinking with interfering radio signals. In our application we have created the interfering noise source by continuously switching a 9V DC supply using a relay kept adjacent to the circuit. The continuous RF interference from relay switches on the LED as seen in Fig.8.

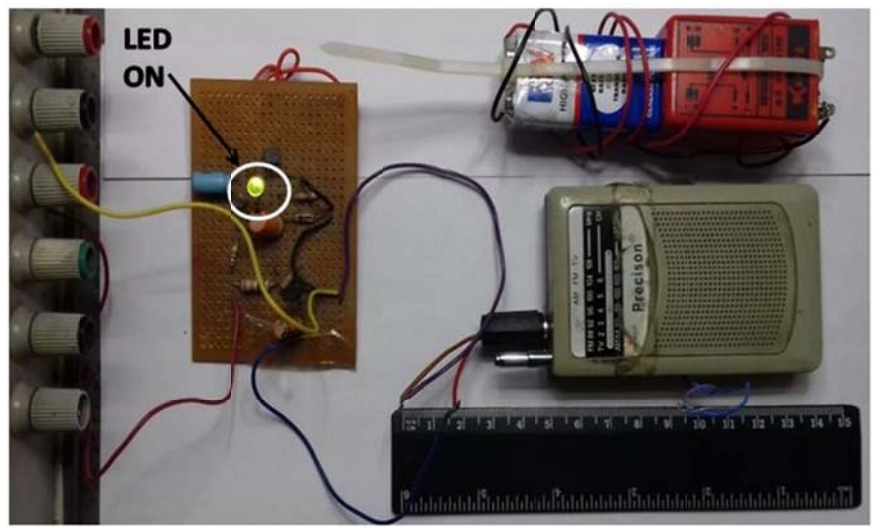

Fig. 8 Radio receiver unit receiving RF interference from a relay oscillator.

This interference from the surrounding sources needs to be masked. For this purpose we have covered the entire relay and battery with our pencil coated papers. We arbitrarily selected six A4 size papers and coated all with pencil strokes. These papers were rolled around relay unit. These rolling of papers added total of 12 layers. As we kept the relay unit inside paper roll the LED blinking disappeared as shown in Fig.9, which indicates effective masking of interference arising from relay.

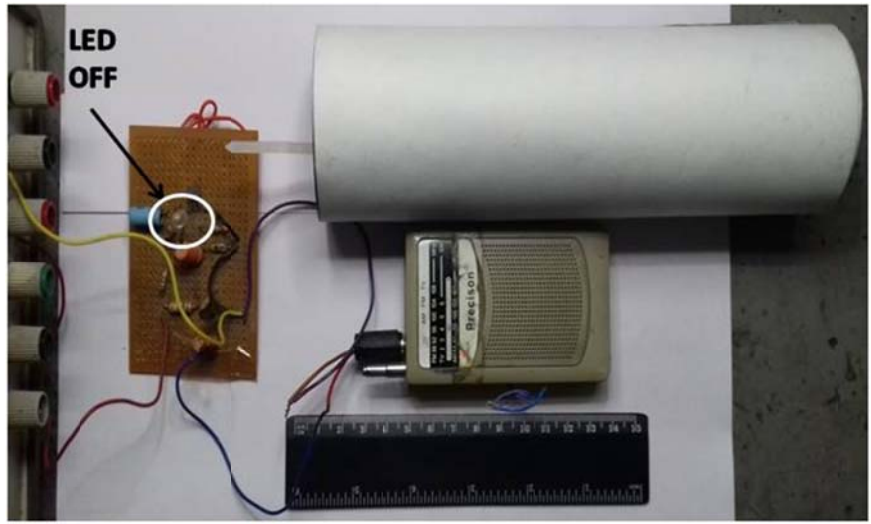

Fig.9. Masking the relay oscillator with pencil coated paper roll.

These observations demonstrate the capability of pencil coated papers to mask the noise signals surrounding any RF device. This indicates the possibility of employing graphite composites from pencil traces as a potential material for RF power attenuating applications.

## 4.0 Conclusion

A very novel method of coating paper with pencil strokes was proposed for studying electromagnetic properties of graphite composites. The graphite composites in the pencil trace provides modulation of the incident RF power. The attenuation characteristic got improved as we go on adding more pencil coated papers onto each other. The increasing thickness of graphite layers helps to enhance the power attenuation capacity for the incident signal. Further, these pencil coated paper roll was able to attenuate the incoming noise signals in the radio signal reception. This demonstrates capability of pencil coated papers to be used for small RF power attenuation applications.

# Supplementary Information

**Monopole Antenna Response**

Monopole antennae were fabricated for five different frequencies in range of 500 MHz-2.5 GHz. Antenna responses was measured by using vector network analyzer. The characteristic spectrum for each antenna is given in following figures. $S_{11}$ measurements were performed for each antenna pair, the observed frequency component and its harmonics is shown in following figures.

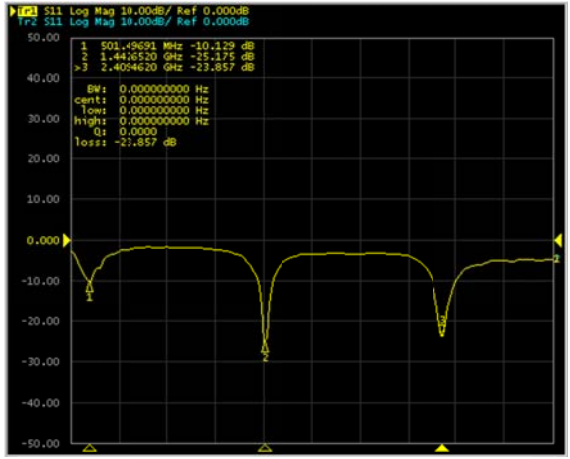 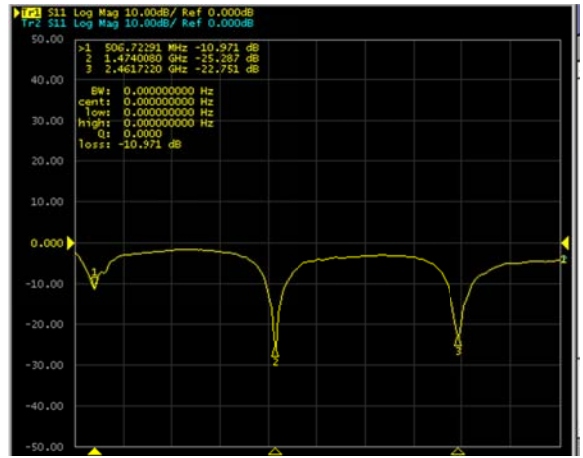

Fig. S1 Antennae spectrum for design frequency of 500 MHz.

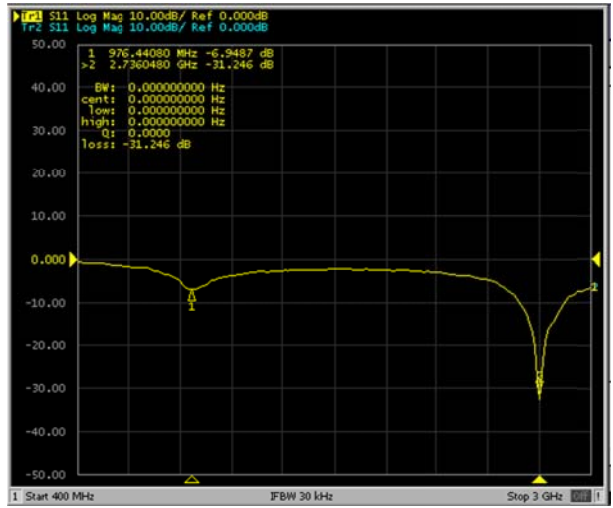 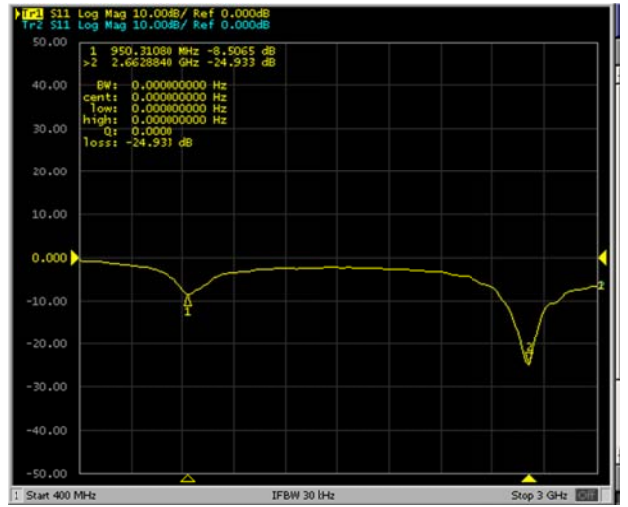

Fig. S2 Antennae spectrum for design frequency of 1GHz.

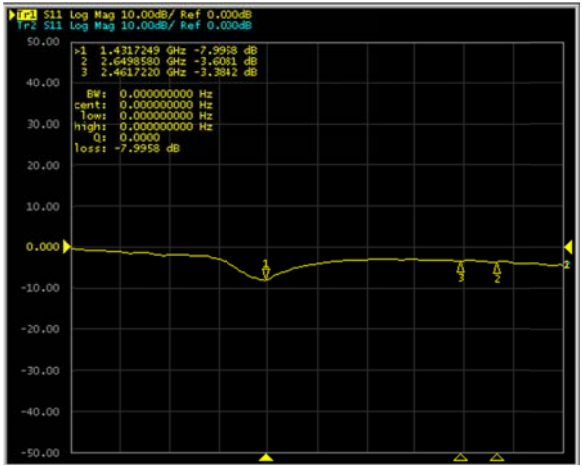 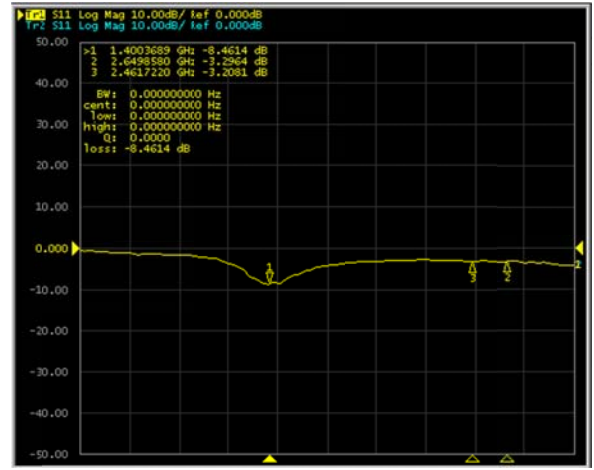

Fig. S3 Antennae spectrum for design frequency of 1.5GHz.

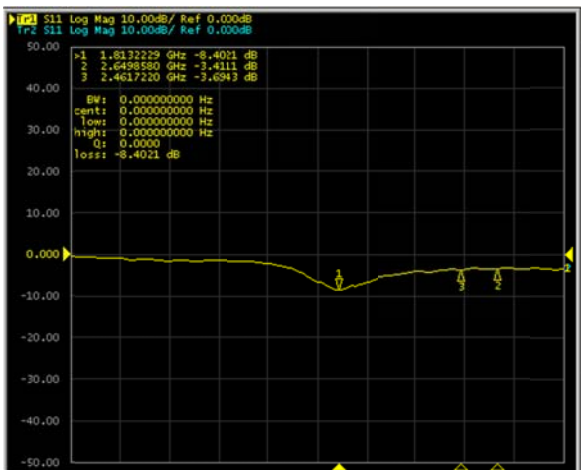 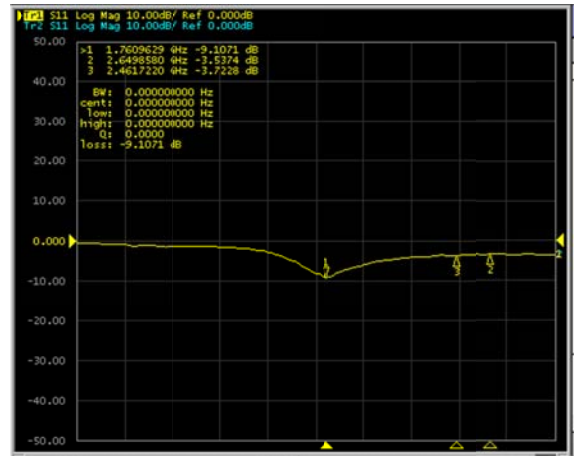

Fig. S4 Antennae spectrum for design frequency of 2 GHz.

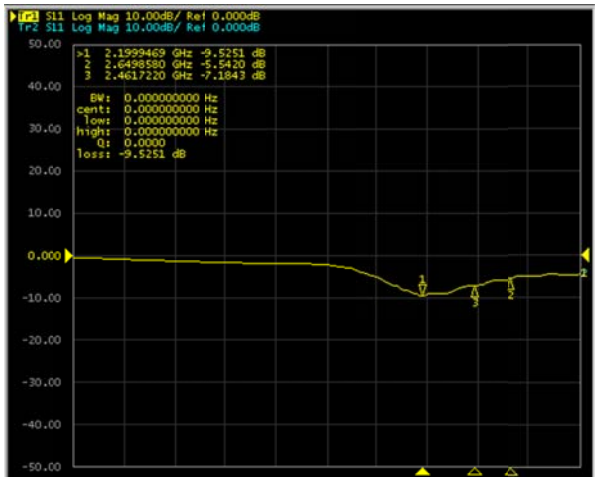 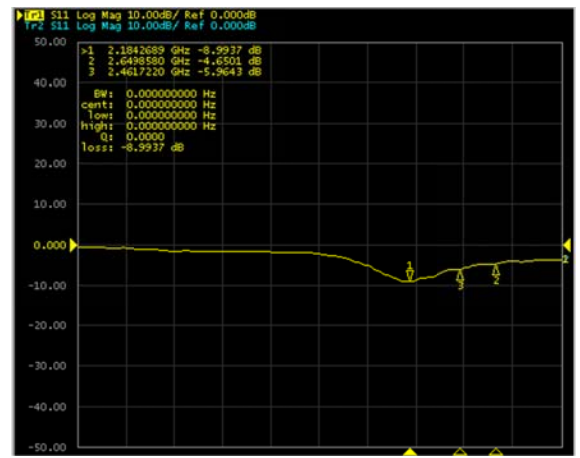

Fig. S5 Antennae spectrum for design frequency of 2.5 GHz.

**Details of Radio control circuit**

The circuit shown in figure S6 is a radio control circuit. That is frequently used to control the remote start a tape recorder, switch on a camera or any other appliance. This circuit basically senses the audio signal input which is further processed, amplified and drives a LED. Once audio signal is received LED at the output will be switched ON. In place of LED relay is mostly used in order to act as control switch for the connected devices.

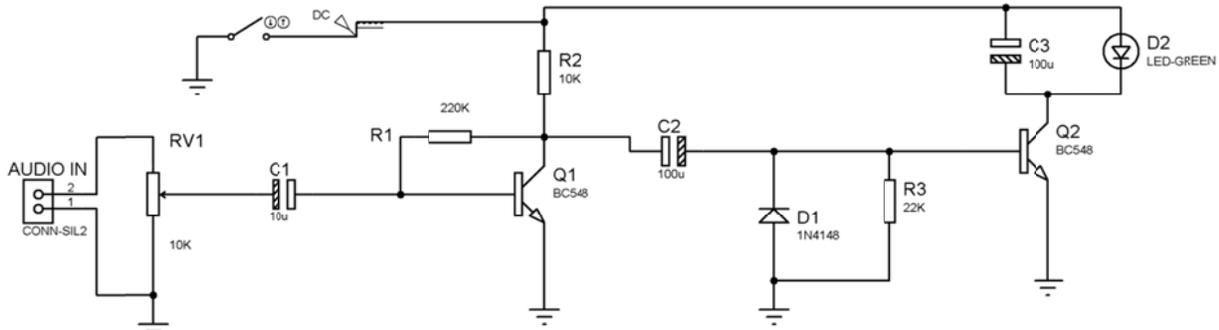

Fig. S6 Schematic of Radio control circuit

In our application we have connected a radio receiver at audio input of the circuit. The radio receiver is set to work AM mode of operation. The receiver is not tuned to any particular AM station frequency so there is no signal received at the input of our control circuit. Therefore LED at the output remains switched OFF. The moment we place a relay oscillator adjacent to control circuit the interference RF signal created from oscillator is sensed by control circuit and subsequently LED switched ON. Now the relay oscillator is covered by pencil coated papers and placed near control circuit. When relay oscillator was switched ON it generated RF signal but due to pencil coated layers radio control circuit was not receiving RF signal and hence LED did not switch ON.